# Up-conversion imaging processing with field-of-view and edge enhancement

Shi-Kai Liu,[1,2] Shi-Long Liu,[1,2] Zhi-Yuan Zhou,[1,2,3,*] Yan Li,[1,2] Yin-Hai Li,[1,2,3] Chen Yang,[1,2] Zhao-Huai Xu,[1,2] Guang-Can Guo,[1,2] and Bao-Sen Shi[1,2,3,*]

[1]*CAS Key Laboratory of Quantum Information, USTC, Hefei, Anhui 230026, China*

[2]*Synergetic Innovation Center of Quantum Information & Quantum Physics, University of Science and Technology of China, Hefei, Anhui 230026, China*

[3]*Wang Da-Heng Collaborative Innovation Center for Science of Quantum Manipulation & Control, Heilongjiang Province & Harbin University of Science and Technology, Harbin 150080, China*

*Corresponding authors: zyzhouphy@ustc.edu.cn; drshi@ustc.edu.cn

Spiral phase contrast is an important and convenient imaging processing technology in edge detection, and a broader field-of-view (FOV) of imaging is a long-pursuing aim to see more regions of the illumination objects. Compared with near-infrared (NIR) spectrum, the up-conversion imaging in visible spectrum benefits from the advantages of higher efficiency detection and lower potential speckle. FOV enhanced and spiral phase contrast up-conversion imaging processing methods by using second order nonlinear frequency up-conversion from NIR spectrum to visible spectrum in two different configurations are presented in this work. By changing the temperature of crystal, controllable spatial patterns of imaging with more than 4.5 times enhancement of FOV is realized in both configurations. Additionally, we present numerical simulations of the phenomenon, which agree well with the experimental observations. Our results provide a very promising way in imaging processing, which may be widely used in biomedicine, remote sensing and up-conversion monitoring.

## Introduction

Optical imaging systems using NIR illumination have a broad application in biomedicine, surveillance and military [1-3]. The wavelength around 1550nm, an eye-safe spectral region, becomes special interests in lidar imaging systems [4]. However, Charge-coupled devices (CCDs) in this spectral domain, commonly based on InGaAs material, suffer from lower efficiency and speed, higher readout noise and more rigorous cooling attachment than CCD in visible spectrum based on silicon. Existing infrared detection limits the technology of direct infrared imaging. So it is necessary to convert the invisible illumination to visible spectrum [5-6]. The up-conversion detection through sum frequency generation (SFG) can visualize infrared image with standard silicon CCD, helping to improve performance and sensitivity of image. For edge and FOV enhancement, two imaging processing methods are introduced respectively below.

Phase contrast can date back to the initial prominent work by Frits Zernike on edge detection in 1930s [7]. He presented a new method for the microscopic observation of transparent objects. In optical imaging system, the spiral phase contrast (SPC) technique has been developed to increase the contrast of intensity and phase objects using a vortex structure filter [8]. In 2005, Jesacher *et al.* realized an oriented shadow effect of SPC imaging of a human cheek cell using a modified spiral phase hologram [9]. Fractional vortex filters were also investigated to obtain gradual edge enhancement by Chen *et al* in 2015[10]. SPC is not just limited on classical region, it can also be translated into the quantum world. In 2009, SPC ghost imaging with correlated photon pairs was proposed by placing a spiral phase plate in one arm, then edge enhancement can occur in another arm [11]. This technique is based on processing the image information in its Fourier plane by placing a spiral phase plate with a topological charge L=1 to serve as a filter. Spiral operation manipulates the phase of the whole light field in the form of a spiral-shaped phase profile of the form $e^{i\varphi}$, where φ is the polar angle in a plane transverse to the light propagation direction [12]. On account of the odd symmetry of spiral phase, both phase and intensity gradients can be enhanced isotropically. Using NIR invisible illumination, SPC imaging visualized by up-conversion may have specific applications on 2D infrared spectroscopy detection. Combining the SPC technology with SFG, both edge enhancement and up-conversion of objective image are obtained simultaneously with nonlinear spatial filter that is periodically poled potassium titanyl phosphate (PPKTP) crystal. This type of trial may pave a fascinating and novel way of pattern recognition and remote sensing in the infrared regime.

Besides, we discover the phenomenon that the FOV of up-conversion imaging varies when the temperature of PPKTP crystal is changing. The traditional methods of increase FOV is realized by using a broad bandwidth pump laser or illumination laser, a dual illumination wavelength, an ASE illumination source, crystal rotation and designing a temperature gradient inside the crystal [13-17]. All of them aim at satisfying the phase matching condition for different incoming signal angles. In contrast to above state-of-the-art approaches, we demonstrate a very convenient and simple method, which has no demand and limits on special required device like laser source or complex temperature gradient controller or precise crystal rotation which causes image intensity and position variation slightly. What we only need to do is changing the temperature of the nonlinear crystal with all the experimental setups remain the same, and more than 4.5 times of FOV enhancement is achieved comparing with the condition of minimum FOV. Besides, temperature scanning is particularly simple and easy way to implement. To the best of our current knowledge, this method has been carried out for the first time and will have wide and useful applications due to the low threshold on the experimental setups and devices. What is worth mentioning is that the same phenomenon of FOV enhancement is found by changing the wavelength of pump laser while we keeps the crystal temperature to be 40.94 ℃, the optimal phase matching temperature, which provides another alternative way to realize this aim. After acquiring up-conversion imaging with temperature of crystal scanned, most illumination part of the object can be seen by "adding" those images together and this can be easily realized by using a time-integration CCD to record image in a range of continuously changing temperature.

## Theory

We demonstrate two different configurations on up-conversion SPC imaging. The generation of SPC image is a linear progress in configuration A, referring to figure 2. Lens L1 and L2 form a 4-f system. $E_{in}(r,\phi)$ is the input object light field, its spatial Fourier spectrum is $E_{in}(\rho,\varphi) = \mathcal{F}[E_{in}(r,\phi)]$ after lens L1. we use the VPP as a spiral phase contrast filter placing on the back spectrum of lens L1 and its transmission function is given by $F(\rho,\varphi) = circ(\frac{\rho}{R})\exp(i\varphi)$, where R is the radius of aperture of $circ(\frac{\rho}{R})$. The whole light field in the Fourier plane is

$$E_f(\rho,\varphi) = F(\rho,\varphi) \times \mathcal{F}[E_{in}(r,\phi)], \quad (1)$$

where × stands for the dot product operation and $\mathcal{F}$ denotes the Fourier transform. Final filtered image in the right focal plane of lens L2 is given by

$$E_{out}(r,\varphi) = \mathcal{F}[F(\rho,\varphi)] * E_{in}(r,\phi), \quad (2)$$

where * represents the convolution. The point spread function is the Fourier transform of the spiral phase filter and it's a doughnut-shaped intensity ring with phase between 0 and $2\pi$ around the ring. With the convolution calculation performed, every single point of the image is weighted with the point spread function and then the integration is carried out all over the illumination part of the resolution target. In "flat" regions, every adjacent point has the constant distribution except at edge where either the phase or the amplitude of adjacent points differ. This process leads to destructive interference in "flat" regions because of $\pi$ out of phase between two points across the center of doughnut [12]. In a similar way, the edge which the gradient is nonzero between regions is bright by constructive interference.

In configuration B, the SPC progress is carried out by PPKTP crystal as a nonlinear filter with an OAM pump beam. That's the main difference compared with configuration A. The up-conversion imaging is visible with an invisible illumination beam by SFG. According to the wave coupling equation, three wave mixing occurs in the Fourier spectrum plane in the crystal, making the two steps which are SPC and up-conversion progress in configuration A becomes one intriguingly. PPKTP crystal is placed at the 2f position of 4-f system and it plays a crucial part of transferring the OAM and object information to the up-conversion image by nonlinear interaction. Vortex filter is placed in the visible pump beam to manipulate the Fourier spectrum information to achieve edge enhancement. We image the vortex filter on the nonlinear crystal. The Fourier transform of the object illuminated by the NIR beam with wavelength $\lambda_s$ is $E_s(\rho,\varphi,\lambda_s) = \mathcal{F}[E_s(r,\phi,\lambda_s)]$, before crystal is put on. The output SFG light field is governed by wave coupling equation (3) and it is simplified by the paraxial approximation and the slowly varying amplitude field approximation.

$$\begin{aligned}
\frac{\partial E_{out}(\lambda_c)}{\partial z} &= \frac{2i\omega_c^2 d_{eff}}{k_c c^2} E_{in}(\lambda_s) F(\lambda_p) e^{-i\Delta kz} + \frac{i}{2k_c}\left(\frac{\partial^2 E_{out}(\lambda_c)}{\partial x^2} + \frac{\partial^2 E_{out}(\lambda_c)}{\partial y^2}\right) \\
\frac{\partial E_{in}(\lambda_s)}{\partial z} &= \frac{2i\omega_s^2 d_{eff}}{k_s c^2} E_{out}(\lambda_c) F(\lambda_p)^* e^{i\Delta kz} + \frac{i}{2k_s}\left(\frac{\partial^2 E_{in}(\lambda_s)}{\partial x^2} + \frac{\partial^2 E_{in}(\lambda_s)}{\partial y^2}\right) \\
\frac{\partial F(\lambda_p)}{\partial z} &= \frac{2i\omega_p^2 d_{eff}}{k_p c^2} E_{out}(\lambda_c) E_{in}(\lambda_s)^* e^{i\Delta kz} + \frac{i}{2k_p}\left(\frac{\partial^2 F(\lambda_p)}{\partial x^2} + \frac{\partial^2 F(\lambda_p)}{\partial y^2}\right)
\end{aligned} \quad . \quad (3)$$

Where $\lambda_c$ is the wavelength of converted SFG beam, $\lambda_p$ is the wavelength of pump beam, $d_{eff}$ is the effective nonlinear coefficient and c is the speed of light in vacuum. The nonlinear term in the equations presents the nonplanar wave distribution of the interaction beams. By imprinting the vortex filter on the nonlinear crystal, the invisible signal image is converted to the visible output SPC image [18].

FOV is a fundamental parameters of imaging systems and how to improve it is a long-sought aim. It's like an adjustable aperture based on the theory of changing the phase matching condition to determine how broad you want to see of the illumination part of objects. For our case, the quasi-phase-matching (QPM) condition is given by $\Delta k = k_{3z} - k_{2z} - k_{1z} - G_m$, where $k$ represents the wave vector. $\Delta k = 0$ means the momentum mismatch is fully compensated by the reciprocal vector $G_m$ of the QPM crystal. By changing the temperature of the PPKTP crystal, the up-conversion controllable spatial pattern with different FOV of SHG beam are obtained and same phenomenon are obtained when we change the wavelength of pump beam or illumination beam. This phenomenon was firstly discovered by Shi's group in 2014 [19]. In Ref. 19, controllable spatial structures of SFG light beam via crystal phase mismatching by tuning the pump frequency or crystal temperature were reported, where two Gaussian pump beams were used. In this work, we "add" imaging processing onto this model and find that the up-convertion image can be tuned periodically, more regions of the illumination object can be seen.

The advantage of up-conversion using QPM compared with angle phase matching is that the largest effective nonlinear coefficient can be used and there is no walk-off effect [20]. Based on the QPM, we can obtain the up-conversion imaging in any arbitrary wavelength by designing the QPM period of crystal and coating differently. Recent work of SPC up-conversion imaging with second harmonic generation using Potassium Titanyl Phosphate (KTP) crystal based on critically type II phase-matching was achieved by Chen's group [18]. It is the first time that the SPC imaging technique has been realized in nonlinear optical process. However, this configuration has the limits of wavelength extendibility, so it can only convert a specific given wavelength of 1064nm with no walk-off effect. To the best of our knowledge, our work is the first time to achieve the SPC up-conversion imaging using illumination beam and pump beam with different wavelength in SFG process and with FOV enhancement.

## Image simulation results

A computational numerical tool is developed to simulate the image processing of edge and FOV enhancement so that we can anticipate the improvements brought by our experimental setup and give a feedback to carry on continuing optimization of our up-conversion system. Fourier spatial filtering method is used to perform the SPC of image. The beam and crystal parameters are in accordance with the experiment. Given the focusing parameters of the pump beams and the phase mismatch caused by temperature change, the spatial structures of the up-conversion image can be simulated numerically using coupled wave functions (3).

We choose an area of USAF-1951 resolution binary image as the input signal. After the image computational simulation processing, the output image is shown in Fig. 1. As we can see, the FOV and the spatial pattern of image changes with the phase mismatching parameter-temperature change. More than 5 times FOV is obtained according to the effective display area theoretically.

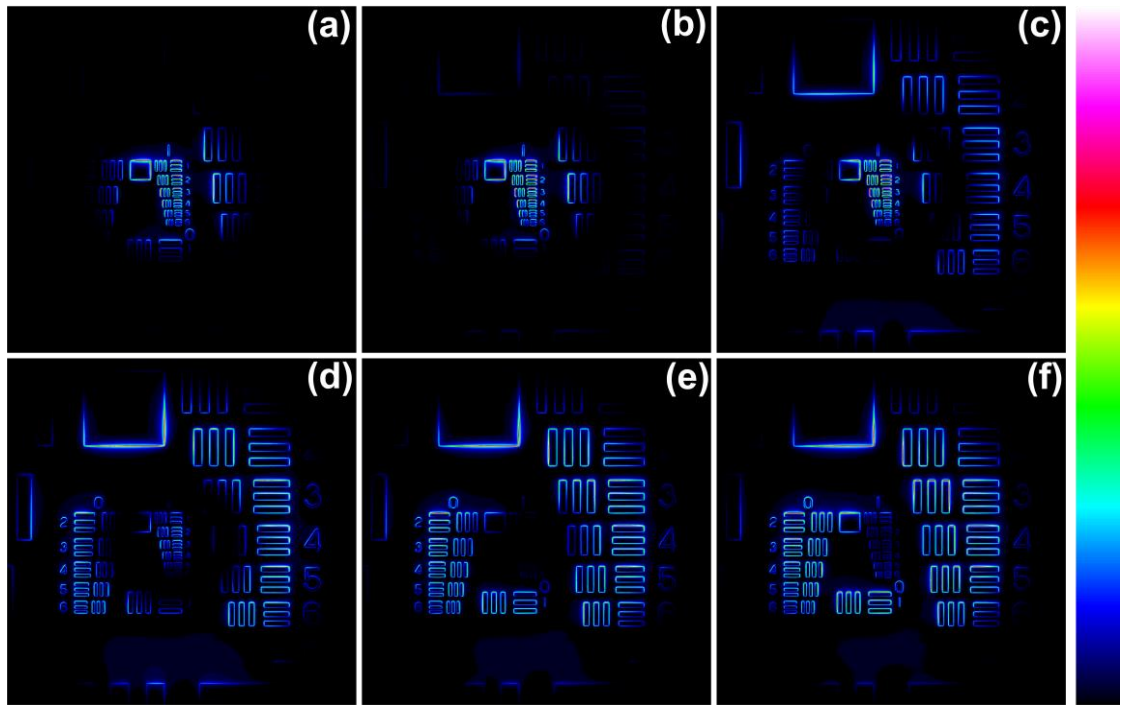

Fig. 1 Image simulation of edge and FOV enhancement of USAF-1951 resolution target. Color bar is given in the right side.

## Experimental setups

Our experimental setup is displayed in Fig. 2. There are two different experimental setup A and B as follows. Firstly the configuration A is introduced, we firstly generate the NIR SPC image, then use a Gaussian pump beam to up-convert this image. By changing the temperature of PPKTP crystal, we acquire the FOV enhanced SPC imaging. The 1556.3nm illumination beam is from a diode laser (Toptica prodesign) and is amplified by an erbium-doped fiber amplifier. The pump beam at 791.8nm wavelength is from Ti:Sapphire laser. Both of the beams are Fiber-coupled by fiber collimators. A USAF-1951 resolution target is inserted as an illumination intensity object. Two pairs of half-wave plates (HWPs) and quarter-wave plates (QWPs) are used in each beam to control the polarization to satisfy the type-I phase matching. The type-I PPKTP crystal we used in our experiment is 3mm long with an aperture of 2mm*1mm with both end faces are anti-reflection coated for 524.8nm, 1556.3nm and 791.8nm. The quasi-phase matching QP is 19.36um. Lens L1, lens L2 and vortex phase plate (VVP) with a topological charge L=1 inserted in the waist of the illumination beam are formed a SPC 4-f imaging system. The topological charge of VPP aligns with the center of the beam. Then lens3 focuses the SPC imaging to PPKTP crystal. The pump beam is transformed into 2 times smaller beams in diameter by the 4-f system consisting of lens L5 and lens L6 whose confocal length is 300mm and 150mm respectively. A dichroic mirror reflects the pump beam and translates the illumination beam to overlap at the same direction. The PPKTP crystal is positioned in the Fourier plane of the up-conversion 4-f system composed of lens L3 and lens L4. The output image was cleaned up by a band pass 525nm filter whose full width at half maximum is 10nm. Then a low-noise and high-speed CCD (BC106-VIS from Thorlabs) records the visible converted imaging.

Another configuration B is basically based on experimental setup A, but with a little change. The VPP @1556nm is removed and insert into the path of pump beam that is the position of the front focal plane of the 4-f system of lens L5 and lens L6. Then we switch the VPP to the wavelength at 792nm and align with the center of pump beam. The image without SPC is generated by the first 4-f system including lens L1 and lens L2. Next the pump beam carrying OAM light mixes with the illumination beam and imprints the VPP on PPKTP crystal during SFG. In this way, the up-conversion and SPC occurs in one step, which is more convenient and simpler.

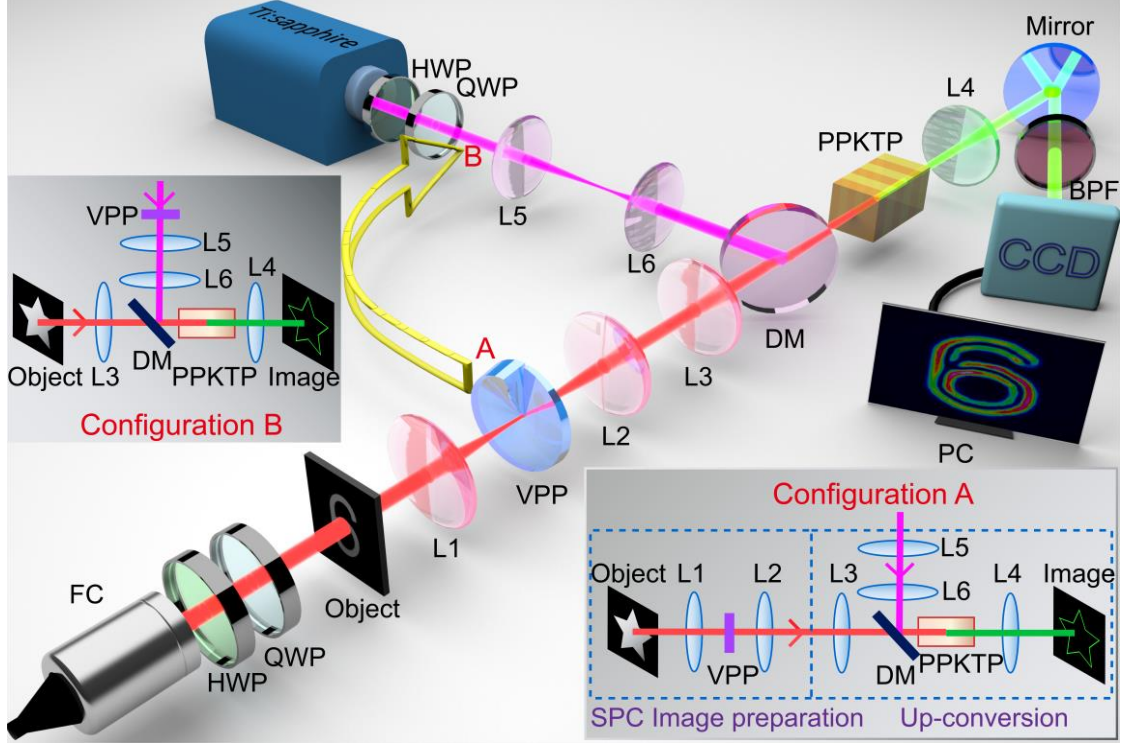

Fig. 2.Experimental setup of SPC up-conversion imaging system containing two configurations. FC, fiber collimator; HWP, half-wave plates; QWP, quarter-wave plates; BPF, band pass filter; DM, dichroic mirror; VPP, vortex phase plate; PPKTP, periodically poled potassium titanyl phosphate crystal; PC, personal computer; L1-L5 are all lens.

## Experimental results

In the above configurations, asymmetric three Arabic numerals 4, 5 and 6 are chosen as the intensity object in order to show the universality of object. Both up-conversion image of single number with or without SPC are all acquired. The output image is shown in Fig. 3. Our experimental results agrees well with the theory and simulation. Through the imaging processing of SPC, we can easily see the outline and the shapely edge enhancement in image (b1)-(b3) and image (c1)-(c3) in Figure 3. Compared with image in configuration B, more speckle and scattered noise emerge in image in another configuration because the NIR illumination beam suffers slightly scattering at the focal plane where VPP is positioned in SPC 4-f system. However，one advantage that configuration A has，but another one lacks，is that the direct NIR image detection can be carried out ,recording SPC image in NIR spectrum using InGaAs CCD [21].

In order to approximately analyze the quality of image, we define the average visibility as the index to evaluate the quality of SPC: $V = \frac{I_{max} - I_{min}}{I_{max} + I_{min}}$. $I_{max}$ is the average maximum gray value of the enhanced outlines, $I_{min}$ is the minimum gray value of dark region between the highlighted outlines. Line chart (d1)-(d3) shows the vertical intensity distribution result from image (a3) to image (c3) according to the red dashed line direction in Fig 3, and the average visibility of those two schemes in image (b3) and image (c3) of number 6 are around 91.8 % and 77.7% respectively.

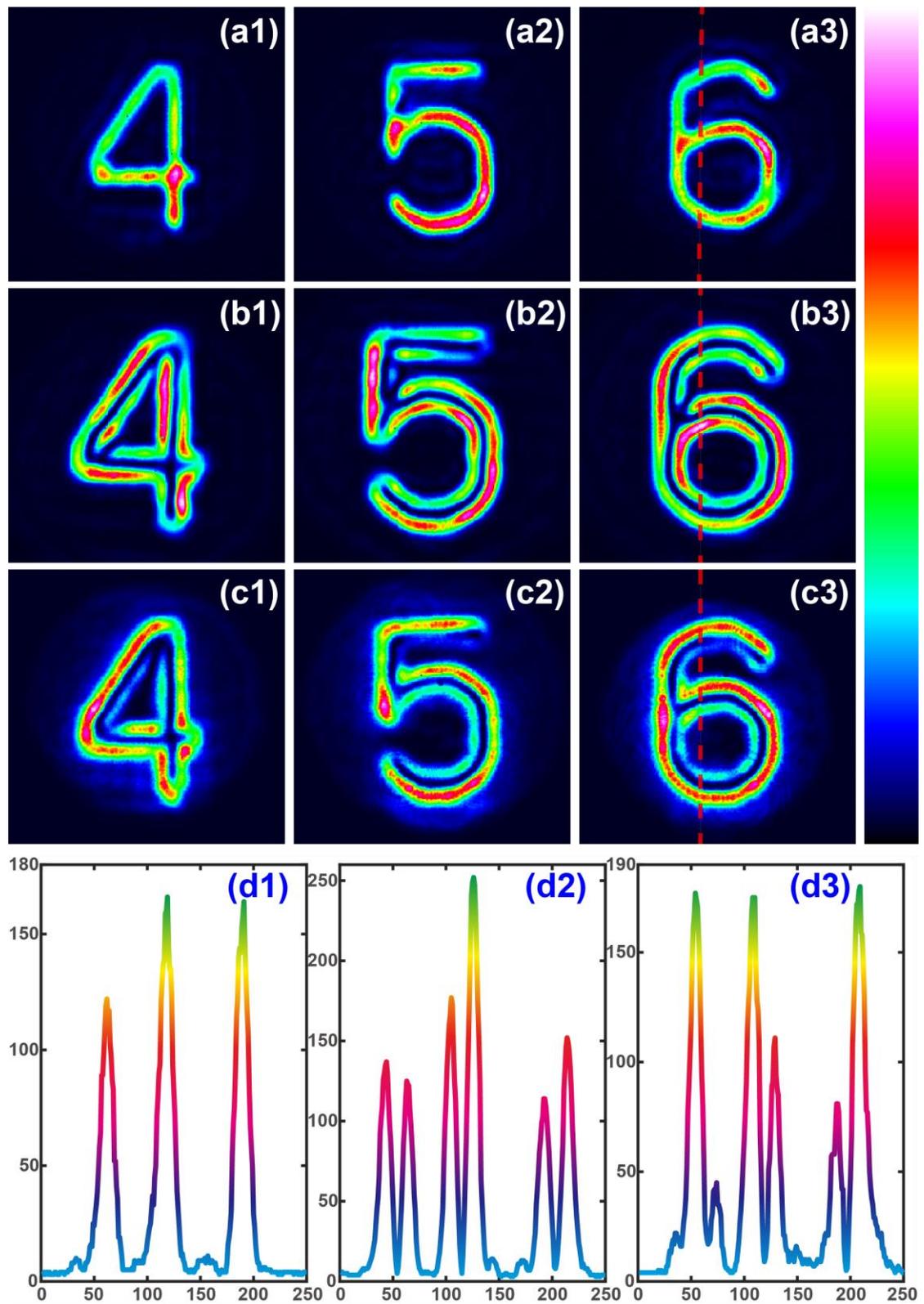

Fig. 3 Experimental up-conversion image of number 4,5 and 6. (a1)-(a3) are image without SPC. (b1)-(b3) are SPC image in configuration B. (c1)-(c3) are SPC image in configuration A. Color

bar is shown in the right side of the figure, all the other figures below use the same color bar so it won't be shown again. Line chart (d1)-(d3) corresponds to the vertical intensity distribution of image (a3)-(c3) respectively.

Another part of USAF-1951 resolution target is chosen to demonstrate the improvement of FOV process. When we adjust the temperature of the PPKTP crystal, different spatial patterns of the converted imaging with different FOV are obtained. The temperature of PPKTP crystal is controlled using a homemade temperature controller with a temperature stability of ±0.002 ℃. The stability of intensity and FOV of image is immune to temperature fluctuation in the range of 0.1 ℃, thus this scheme can be easily realized in the practical life and complicated circumstances due to the high temperature tolerance and stabilization. Obviously, the FOV in different temperature situation varies due to the phase mismatching, demonstrated in Fig 4. By calculating the area of the bright region, we get over 4.5 times FOV enhancement in comparison with image (b) and image (d). This interesting process is much more like an "adjustable aperture" by analogy with different wide-angle lens of camera. Also, the acutance and contrast ratio still remains in the wide FOV situation such as image (d).

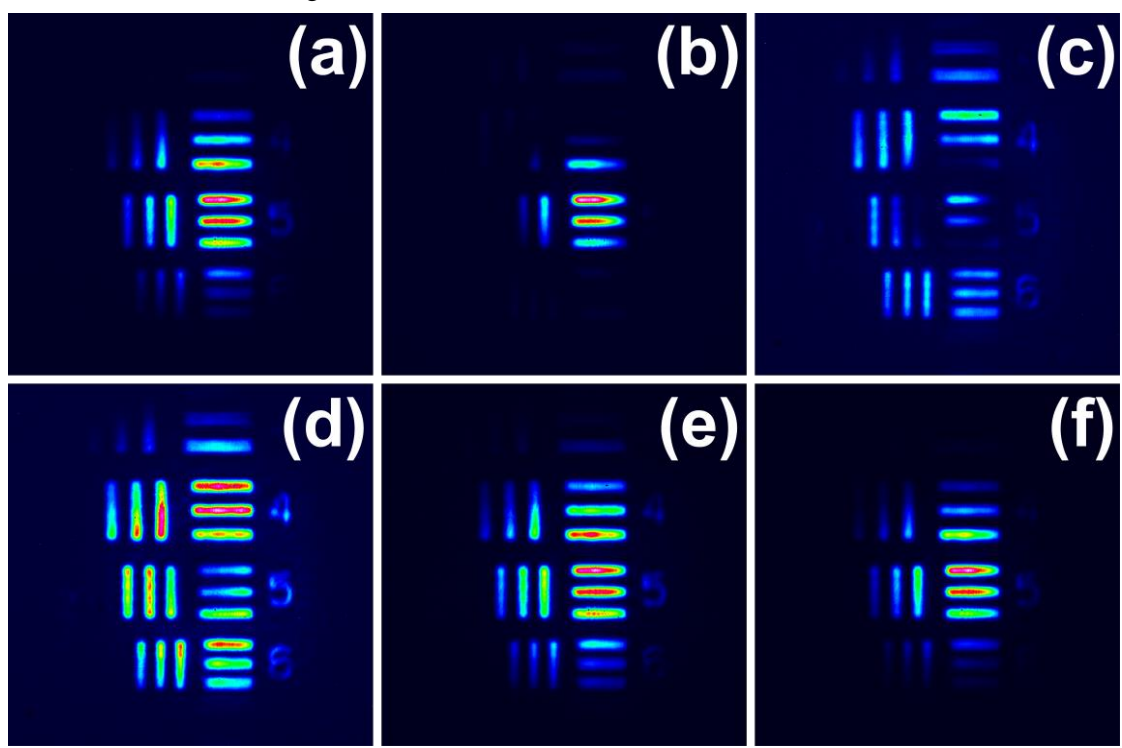

Fig. 4 The temperature controllable FOV of image are shown in (a)-(f). The temperature condition in image (a)-(f) is 18.7 ℃, 29.6 ℃, 33.0 ℃, 34.5 ℃, 37.1 ℃ and 45.7 ℃ respectively.

Next we discuss the imaging processing of both FOV and edge enhancement at the same time in two configurations. Figure. 5 depicts all the outcomes of up-conversion image. Two different places of USAF-1951 resolution target are used as illumination part in those two configurations. As is vividly presented in the image, the outline of the whole intensity object is highlighted with different spatial pattern by controlling PPKTP crystal's temperature. FOV is maximally enhanced in image (a4) and image (b4) and we acquire more than 4.5 times FOV enhancement compared with the image in minimal FOV. Image scanning of the illumination part of object happens with the crystal's temperature changing and the quality of the above image remains the same with the temperature changing. Image in configuration A has slight stray light in the background due to the diffraction of VPP positioned at focal plane. In general, those two configurations are basically equivalent in theory and experiment, only depending on which light beam arm carrying OAM.

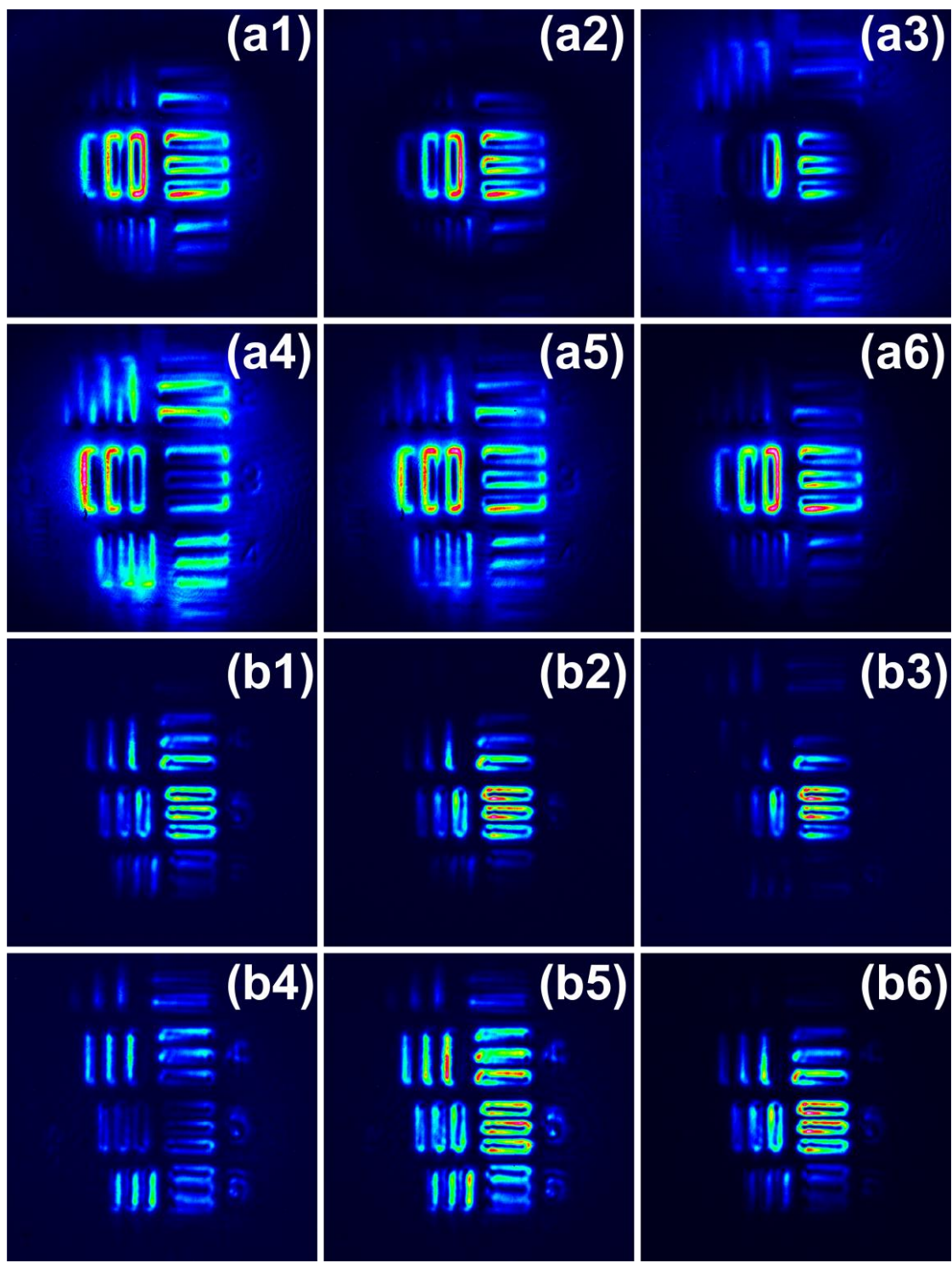

Fig. 5 The up-conversion SPC image in different temperature. Image (a1)-(a6) is from configuration A and image (b1)-(b6) is from configuration B. The temperature condition from image (a1)-(a6) or image (b1)-(b6) is 18.7 ℃, 24.1 ℃, 29.6 ℃, 34.5 ℃, 37.1 ℃, and 45.7 ℃ respectively.

## Conclusion

In summary, we demonstrate two imaging processing methods with both FOV and edge enhancement separately and simultaneously via up-conversion using invisible illumination in two configurations. Since the optical method has the advantage of parallel processing and high-speed, it can also apply on real-time holographic display via spiral light modulator [18, 22]. Up-conversion SPC imaging will have application on reagent-free biological imaging and infrared monitoring in the next future. In addition, by temperature control or wavelength changing, image with different FOV enhance also obtained. To forecast, the flexibility and extensibility of our experimental up-conversion setup will provide several potential choices to process image based on multiple specific circumstance. The efficiency of up-conversion can be improved by using intra-cavity or using high intensity pulsed light [6]. Also, this system may be performed on single-photon up-conversion ghost imaging using correlated photon pairs after modification of experimental setup, paving the way to quantum imaging processing.

## Acknowledgments

This work is supported by the National Natural Science Foundation of China (NSFC) (61435011, 61525504, 61605194); Anhui Initiative In Quantum Information Technologies (AHY020200); the China Postdoctoral Science Foundation (2016M590570); and the Fundamental Research Funds for the Central Universities.

## References

1.  Uday K. Tirlapur and Karsten König, "Cell biology: Targeted transfection by femtosecond laser," Nature 418, pages 290–291 (2002)
2.  S. Gioux, HS. Choi, and JV. Frangioni, "Image-Guided Surgery Using Invisible Near-Infrared Light: Fundamentals of Clinical Translation," Mol. Imaging Vol 9, No 5, 7290-2010 (2010)
3.  R. Hudson and J. Hudson, "The military applications of remote sensing by infrared," Proc. IEEE 63, 104–128 (1975).
4.  C.-L. Wang, X.-D. Mei, L Pan, P.-W Wang, W Li, X-Gao, Z.-W. Bo, M.-L.Chen,W.-L. Gong and S.-S. Han, "Airborne Near Infrared Three-Dimensional Ghost Imaging LiDAR via Sparsity Constraint," Remote Sensing, 10(5), 732(2018).
5.  A. P. Vandevender and P. G. Kwiat, "High efficiency single photon detection via frequency upconversion," J. Mod. Opt. 51, 1433–1445(2004).
6.  A. J. Torregrosa, H. Maestre, and J. Capmany, "Intra-cavity upconversion to 631 nm of image illuminated by an eye-safe ASE source at 1550 nm, " Opt. Lett. Vol. 40, No. 22, 5315-5318 (2015).
7.  F. Zernike, "Phase contrast, a new method for the microscopic observation of transparent objects," Physica 9, 686–698 (1942).
8.  S. Fürhapter, A. Jesacher, S. Bernet, and M. Ritsch-Marte, "Spiral phase contrast imaging in microscopy," Opt. Express 13, 689–694 (2005).
9.  Jesacher A, Fürhapter S, Bernet S, and Ritsch-MarteM, "Shadow effects in spiral phase contrast microscopy," Phys. Rev. Lett. 94, 233902 (2005).
10. J.-K. Wang, W.-H. Zhang, Q.-Q. Qi, S.-S. Zheng, and L.-X. Chen, "Gradual edge enhancement in spiral phase contrast imaging with fractional vortex filters" , Sci. Rep. 5, 15826 (2015).
11. B. Jack, J. Leach, J. Romero, S. Franke-Arnold, M. Ritsch-Marte, S. M. Barnett, and M. J.

Padgett, "Holographic Ghost Imaging and the Violation of a Bell Inequality," Phys. Rev. Lett. 103, 083602 (2009).
12. M. Ritsch-Marte, "Orbital angular momentum light in microscopy," Philos.Trans. R. Soc. London A 375, 20150437 (2017).
13. S. Junaid, J. Tomko, M.P. Semtsiv, J. Kischkat, W.T. Masselink, C. Pedersen, and P. Tidemand-Lichtenberg, "Mid-infrared upconversion based hyperspectral imaging," Opt. Express, 26(3), 2203-2211 (2018).
14. R. Demur, R. Garioud, A. Grisard, E. Lallier, L. Leviandier, L. Morvan, N. Treps, and C. Fabre, "Near-infrared to visible upconversion imaging using a broadband pump laser, " Opt. Express, 26(10), 13252-13263 (2018).
15. H. Maestre, A. J. Torregrosa, and J. Capmany, "IR image upconversion under dual-wavelength laser illumination," IEEE Photonics J. 8(6), 1-8 (2016).
16. A. Torregrosa, H. Maestre, and J. Capmany, "Intra-cavity upconversion to 631 nm of image illuminated by an eye-safe ASE source at 1550 nm," Opt. Lett. 40(22), 5315–5318 (2015).
17. H. Maestre, A. J. Torregrosa, C. R. Fernández-Pousa, and J. Capmany, "IR-to-visible image upconverter under nonlinear crystal thermal gradient operation," Opt. Express. 26(2), 1133-1144 (2018).
18. X.-D. Qiu, F.-S. Li, W.-H. Zhang, Z.-H. Zhu, and L.-X. Chen, "Spiral phase contrast imaging in nonlinear optics: seeing phase objects using invisible illumination," Optica, 5(2), 208-212(2018).
19. Z.-Y. Zhou, Y Li, D.-S. Ding, Y.-K. Jiang, W Zhang, S Shi, B.-S. Shi and G.-C. Guo, "Generation of light with controllable spatial patterns via the sum frequency in quasi-phase matching crystals," Sci. Rep. 4, 5650 (2014).
20. M. Pierrou, F. Laurell, H. Karlsson, T. Kellner, C. Czeranowsky, and G. Huber, "Generation of 740 mW of blue light by intracavity frequency doubling with a first-order quasi-phase-matched KTiOPO4 crystal," Opt. Lett. 24(4), 205-207 (1999).
21. Marshall J. Cohen, Gregory H. Olsen, "Room-temperature InGaAs camera for NIR imaging", Proc. SPIE 1946, Infrared Detectors and Instrumentation, (1993).
22. F. Yaraş, H. Kang, and L. Onural. "Real-time phase-only color holographic video display system using LED illumination," Appl. Opt. 48.34: H48-H53 (2009).